\documentclass[aps,prd,onecolumn,groupedaddress,showpacs,nofootinbib,amssymb]{revtex4}
\usepackage[dvips]{graphicx}
\usepackage{amssymb}
\usepackage{amsmath}
\usepackage{graphicx,,color}
\usepackage{amsfonts}
\usepackage{bm}
\usepackage{cancel}
\usepackage{comment}

\newcommand\be{\begin{equation}}
\newcommand\ee{\end{equation}}

\allowdisplaybreaks[4]

\begin{document}

\tolerance=5000

\title{Canonical Scalar Field Inflation with String and $R^2$-Corrections}
\author{S.D.~Odintsov,$^{1,2}$\,\thanks{odintsov@ieec.uab.es}
V.K.~Oikonomou,$^{3,4,5}$\,\thanks{v.k.oikonomou1979@gmail.com} F.P.
Fronimos,$^{3}$\,\thanks{fotisfronimos@gmail.com}}
\affiliation{$^{1)}$ ICREA, Passeig Luis Companys, 23, 08010 Barcelona, Spain\\
$^{2)}$ Institute of Space Sciences (IEEC-CSIC) C. Can Magrans
s/n,
08193 Barcelona, Spain\\
$^{3)}$ Department of Physics, Aristotle University of
Thessaloniki, Thessaloniki 54124,
Greece\\
$^{4)}$ Laboratory for Theoretical Cosmology, Tomsk State
University of Control Systems and Radioelectronics, 634050 Tomsk,
Russia (TUSUR)\\
$^{5)}$ Tomsk State Pedagogical University, 634061 Tomsk,
Russia\\}

\tolerance=5000

\begin{abstract}
Assuming that a scalar field controls the inflationary era, we
examine the combined effects of string and $f(R)$ gravity
corrections on the inflationary dynamics of canonical scalar field
inflation, imposing the constraint that the speed of the
primordial gravitational waves is equal to that of light's.
Particularly, we study the inflationary dynamics of an
Einstein-Gauss-Bonnet gravity in the presence of $\alpha R^2$
corrections, where $\alpha$ is a free coupling parameter. As it
was the case in the pure Einstein-Gauss-Bonnet gravity, the
realization that the gravitational waves propagate through
spacetime with the velocity of light, imposes the constraint that
the Gauss-Bonnet coupling function $\xi(\phi)$ obeys the
differential equation $\ddot\xi=H\dot\xi$, where $H$ is the Hubble
rate. Subsequently, a relation for the time derivative of the
scalar field is extracted which implies that the scalar functions
of the model, which are the Gauss-Bonnet coupling and the scalar
potential, are interconnected and simply designating one of them
specifies the other immediately. In this framework, it is useful
to freely designate $\xi(\phi)$ and extract the corresponding
scalar potential from the equations of motion but the opposite is
still feasible. We demonstrate that the model can produce a viable
inflationary phenomenology and for a wide range of the free
parameters. Also, a mentionable issue is that when the coupling
parameter $\alpha$ of the $R^2$ correction term is
$\alpha<10^{-3}$ in Planck Units, the $R^2$ term is practically
negligible and one obtains the same equations of motion as in the
pure Einstein-Gauss-Bonnet theory, however the dynamics still
change, since now the time derivative of $\frac{\partial
f}{\partial R}$ is nonzero. We study in detail the dynamics of
inflation assuming that the slow-roll conditions hold true and
also we briefly address the constant-roll case dynamics, and by
using several illustrative examples, we compare the dynamics of
the pure and $R^2$-corrected Einstein-Gauss-Bonnet gravity.
Finally the ghosts issue is briefly addressed.
\end{abstract}


\maketitle

\section{Introduction}

One good question in theoretical physics has always been if the
constants of nature are actually constant, or have been
dynamically evolving to their value we actually estimate at
present time. Tough to answer actually, but it is our opinion that
during and possible after the inflationary era, the Universe had
settled to a classical state, so all the particles received their
mass, and all constants of nature that do not depend inherently on
the comoving radius of the Universe, received the value that we
measured at present time.

In this line of research, the striking GW170817 event
\cite{GBM:2017lvd} involved a Kilonova which indicated that the
gravitational wave speed $c_T$ is equal to that of light, so
$c_T^2=1$ in natural units. Now the question is whether it is
compelling to assume that the speed of the primordial
gravitational waves is also equal to that of light's. To our
opinion it is since the constants of nature during the
inflationary and the post-inflationary era, must actually be
constants. This way of thinking is also adopted in the literature,
so as was shown in Ref. \cite{Ezquiaga:2017ekz}, many theoretical
frameworks were actually put into serious question after the
striking GW170817 event.

As we already mentioned previously, the inflationary era is
basically a classical era or at least semi-classical, in the sense
that the Universe is already four dimensional and expands and
simultaneously cools significantly. Since the inflationary era
emerged from a quantum era, it is natural to assume that some
imprints of the quantum era may be present in the effective
Lagrangian that describes the inflationary era. Since string
theory in its various aspects is to date the most prominent
candidate theory able to describe the quantum era, it is natural
to assume that string corrections may actually be present in the
effective Lagrangian that describes the inflationary era. A well
studied class of string-corrected gravitational theories, is known
under the name Einstein-Gauss-Bonnet theory, and it is quite
popular in inflationary and astrophysical studies
\cite{Hwang:2005hb,Nojiri:2006je,Cognola:2006sp,Nojiri:2005vv,Nojiri:2005jg,Satoh:2007gn,Bamba:2014zoa,Yi:2018gse,Guo:2009uk,Guo:2010jr,Jiang:2013gza,Kanti:2015pda,vandeBruck:2017voa,Kanti:1998jd,Pozdeeva:2020apf,Fomin:2020hfh,DeLaurentis:2015fea,Chervon:2019sey,Nozari:2017rta,Odintsov:2018zhw,Kawai:1998ab,Yi:2018dhl,vandeBruck:2016xvt,Kleihaus:2019rbg,Bakopoulos:2019tvc,Maeda:2011zn,Bakopoulos:2020dfg,Ai:2020peo,Odintsov:2019clh,Oikonomou:2020oil,Odintsov:2020xji,Oikonomou:2020sij,Odintsov:2020zkl,Odintsov:2020sqy,Easther:1996yd,Antoniadis:1993jc,Antoniadis:1990uu,Kanti:1995vq,Kanti:1997br}.
The Einstein-Gauss-Bonnet theory was crucially affected by the
GW170817 results, since the primordial gravitational wave speed
for this theory is not equal to unity in natural units, but it is
equal to $c_T^2=1-\frac{Q_f}{2Q_t}$ where
$Q_f=8c_1(\ddot\xi-H\dot\xi)$. Recently, we pointed out in several
articles that if we manage to make the term $Q_f$ equal to zero,
the speed of the primordial gravitational waves can be equal to
unity in natural units, for the Einstein-Gauss-Bonnet
\cite{Odintsov:2019clh,Oikonomou:2020oil,Odintsov:2020xji,Odintsov:2020zkl,Odintsov:2020sqy}
and related theories \cite{Oikonomou:2020sij}, and this constraint
actually relates directly the scalar field potential $V(\phi)$ and
the Gauss-Bonnet scalar coupling function $\xi(\phi)$. In the same
spirit in this work we consider the effects of $R^2$-corrections
of the inflationary phenomenology of the Einstein-Gauss-Bonnet
theory. The motivation for this is mainly the fact that modified
gravity theories
\cite{Nojiri:2017ncd,Nojiri:2010wj,Nojiri:2006ri,Capozziello:2011et,Capozziello:2010zz,Olmo:2011uz},
and particularly, $f(R)$ gravity, is very successful in describing
several evolutionary eras of our Universe, such as the late-time
and early-time era, and sometimes can also provide a unified
description of the acceleration eras of the Universe see for
example Ref. \cite{Nojiri:2003ft} and also Ref. \cite{Sa:2020qfd}
for a recent article in the context of Palatini gravity. Thus it
is interesting to see how such $f(R)$ gravity terms may affect the
inflationary phenomenology of the string-corrected scalar field
Lagrangian. Basically, it is of the same spirit as the
Gauss-Bonnet correction is in the scalar field Lagrangian. The
central premise of this article is thus, that the scalar field
basically controls the inflationary era, and the scalar field
Lagrangian is corrected by string terms and $f(R)$ gravity terms.
We study in detail the inflationary phenomenology of the
$R^2$-corrected theory in both the slow-roll and constant-roll
cases, and we compare the theory with the latest Planck (2018)
observational data \cite{Akrami:2018odb}.

\section{Theoretical Framework of $R^2$ Gravity With String corrections}

The general expression of the string and $f(R)$ gravity corrected
canonical scalar field gravitational action is the following,
\begin{equation}
\centering
\label{action}
S=\int{d^4x\sqrt{-g}\left(\frac{f(R)}{2\kappa^2}-\frac{1}{2}\omega g^{\mu\nu}\partial_\mu\phi\partial_\nu\phi-V(\phi)-\xi(\phi)\mathcal{G}\right)}\, ,
\end{equation}
where $R$ denotes the Ricci scalar, $\kappa=\frac{1}{M_P}$ with
$M_P$ being the reduced Planck mass, $V(\phi)$ the scalar
potential and $\xi(\phi)$ the Gauss-Bonnet coupling scalar
function. Also for the purposes of this paper we shall assume that
$f(R)=R+\alpha R^2$, where $\alpha$ is a constant with mass
dimensions $[m]^{-2}$, but we keep the $f(R)$ function notation in
order to provide general expressions for the equations of motion,
so for the sake of generality. Additionally, $\mathcal{G}$ denotes
the Gauss-Bonnet topological invariant defined as
$\mathcal{G}=R_{\mu\nu\sigma\rho}R^{\mu\nu\sigma\rho}-4R_{\mu\nu}R^{\mu\nu}+R^2$
with $R_{\mu\nu\sigma\rho}$ and $R_{\mu\nu}$ being the Riemann
curvature tensor and the Ricci tensor respectively. Furthermore,
concerning the metric, we shall assume it is a flat
Friedman-Robertson-Walker hence the line element read,
\begin{equation}
\centering
\label{metric}
ds^2=-dt^2+a^2(t)\sum_{i=1}^{3}{(dx^i)^2}\, ,
\end{equation}
where $a(t)$ denotes as usually the scale factor. Consequently,
since the cosmological background is flat, the Ricci scalar and
the Gauss-Bonnet topological invariant are written in terms of
Hubble's parameter $H=\frac{\dot a}{a}$ as $R=6(2H^2+\dot H)$ and
$\mathcal{G}=24H^2(\dot H+H^2)$, where the ``dot'' as usual
implies differentiation with respect to cosmic time $t$. Also,
$\omega$ shall be equated to unity in order to derive the
description of the canonical case, but it shall not be replaced
with unity in order to keep general expressions for the subsequent
equations containing $\omega$, and thus having the phantom case
corresponding to $\omega=-1$ available as well for the interested
reader.

Since string corrections are implemented, gravitational waves
propagate through spacetime with a velocity which does not
necessarily coincide with the speed of light. Specifically, the
speed of the primordial tensor perturbations is in our case equal
to
\cite{Hwang:2005hb,Oikonomou:2020oil,Odintsov:2020xji,Oikonomou:2020sij,Odintsov:2020sqy,Odintsov:2020zkl},
\begin{equation}
\centering
\label{cT}
c_T^2=1-\frac{Q_f}{2Q_t}\, ,
\end{equation}
in natural units and in addition, $Q_f=16(\ddot\xi-H\dot\xi)$ and
$Q_t=\frac{F}{\kappa^2}-8\dot\xi H$ with $F=\frac{\partial
f}{\partial R}$. As a result, compatibility with the GW170817
event can be restored by demanding that the Gauss-Bonnet coupling
satisfies the differential equation,
\begin{equation}\label{diffeqn}
\ddot\xi=H\dot\xi\, ,
\end{equation}
that $Q_f=0$. As was shown in Refs.
\cite{Oikonomou:2020oil,Odintsov:2020xji,Oikonomou:2020sij,Odintsov:2020sqy,Odintsov:2020zkl},
rewriting the differential equation (\ref{diffeqn}) in terms of
the scalar field and assuming that the slow-roll conditions hold
true, meaning that $\ddot\phi\ll H\dot\phi$ for the scalar field,
then we get,
\begin{equation}
\centering
\label{dotphi}
\dot\phi=H\frac{\xi'}{\xi''}\, ,
\end{equation}
where ``prime'' denotes differentiation with respect to the scalar
field $\phi$. By varying the action (\ref{action}) with respect to
the metric tensor $g^{\mu\nu}$ and the scalar field $\phi$, the
equations of motion are obtained which in this case read,
\begin{equation}
\centering
\label{motion1}
\frac{3FH^2}{\kappa^2}=\frac{1}{2}\omega\dot\phi^2+V+\frac{FR-f}{2\kappa^2}-\frac{3H\dot F}{\kappa^2}+24\dot\xi H^3\, ,
\end{equation}
\begin{equation}
\centering
\label{motion2}
-\frac{2F\dot H}{\kappa^2}=\omega\dot\phi^2-16\dot\xi H\dot H+\frac{\ddot F-H\dot F}{\kappa^2}\, ,
\end{equation}
\begin{equation}
\centering \label{motion3}
V'+\omega(\ddot\phi+3H\dot\phi)+\xi'\mathcal{G}=0\, .
\end{equation}
In this case, we shall implement only the slow-roll conditions,
\begin{align}
\centering \label{slowroll} \dot H&\ll H^2\,
,&\frac{1}{2}\omega\dot\phi^2&\ll V\, ,&\ddot\phi&\ll H\dot\phi\,
,
\end{align}
thus, the equations of motion can be simplified greatly. Recalling
that $f(R)=R+\alpha R^2$, one obtains elegant simplifications and
functional expressions. For instance, since $f(R)$ has this
specific form, we have $3FH^2=3H^2+\frac{FR-f}{2}$ and in
addition, $2F\dot H=2\dot H+H\dot F$. Furthermore, $\dot F\sim\dot
{H} H$ hence it can be neglected from the first equation of motion
due to the slow-roll conditions. Hence, the first two equations of
motion are written as,
\begin{equation}
\centering
\label{motion1a}
\frac{3H^2}{\kappa^2}=V-\frac{144\alpha H^2\dot H}{\kappa^2}+24\dot\xi H^3\, ,
\end{equation}
\begin{equation}
\centering \label{motion2a} -\frac{2\dot
H}{\kappa^2}=\omega\dot\phi^2+\frac{\ddot F}{\kappa^2}-16\dot\xi
H\dot H\,.
\end{equation}
Here, we need to note that for $\dot H\ll H^2$, the Ricci scalar
becomes approximately equal to $R\simeq 12 H^2$ and therefore
$F=1+24\alpha H^2$, $\dot F=48\alpha H\dot H$ and $\ddot
F=48\alpha(\dot H^2+H\ddot H)$. Naturally, from the slow-roll
approximations, it stands to reason that $\dot F$ and $\ddot F$
are relative small. In the following, we shall neglect their
contribution from the equations of motion and afterwards ascertain
whether such assumption is in fact reasonable.  Therefore,
assuming that $\ddot F\ll \omega\dot\phi^2$ and discarding the
term $144\alpha H^2\dot H$, one obtains the same equations of
motion as in the simple Einstein-Gauss-Bonnet case, as shown
below,
\begin{equation}
\centering
\label{motion4}
H^2\simeq\frac{\kappa^2V}{3}\, ,
\end{equation}
\begin{equation}
\centering
\label{motion5}
\dot H\simeq-H^2\frac{\kappa^2\omega}{2}\left(\frac{\xi'}{\xi''}\right)^2\, ,
\end{equation}
\begin{equation}
\centering
\label{motion6}
V'+3\omega H\dot\phi+24\xi'H^4\simeq0\, ,
\end{equation}
Here, we mention again that two assumptions were made. Firstly,
the slow-roll conditions, from which the derivatives of the Ricci
scalar have been neglected and secondly, the string corrections
are negligible in the equations of motion. As a result, Hubble's
parameter is given by a simple expression which as a matter of
fact coincides with the case of $\alpha\to 0$ along with Hubble's
derivative and the continuity equation for the scalar field, which
in this framework serves as a differential equation satisfied by
the scalar potential. Thus, the squared term $\alpha R^2$
participates only indirectly in the inflationary phenomenology,
via the expressions of the slow-roll indices and the corresponding
observational indices. In order to see this at first hand, we
shall examine the same models as in Ref. \cite{Odintsov:2020sqy},
since these scalar coupling functions lead to simple expressions
for the ratio $\xi'/\xi''$. It is expected however that a change
will occur since now $\dot F\neq0$, hence the third slow-roll
index shall be present. For $f(R)=R+\alpha R^\gamma$, one obtains,
\begin{equation}
\centering \label{fexample}
\frac{3FH^2}{\kappa^2}+\frac{f-FR}{2\kappa^2}=\frac{3H^2}{\kappa^2}+\frac{\alpha
(12H^2)^\gamma}{\kappa^2}\left(\frac{\gamma}{4}-\frac{\gamma-1}{2}\right)\,
.
\end{equation}
The second term is simplified greatly for $\gamma=2$, hence the
reason $f(R)=R+\alpha R^2$ was selected. Since the slow-roll
conditions are implemented, or at least are assumed to hold true,
the slow-roll indices are defined as follows
\cite{Hwang:2005hb,Oikonomou:2020oil,Odintsov:2020xji,Oikonomou:2020sij,Odintsov:2020sqy,Odintsov:2020zkl},
\begin{align}
\centering \label{indices} \epsilon_1&=-\frac{\dot
H}{H^2}&\epsilon_2&=\frac{\ddot\phi}{H\dot\phi}&\epsilon_3&=\frac{\dot
F}{2HF}&\epsilon_4&=\frac{\dot E}{2HE}&\epsilon_5&=\frac{\dot
F+\kappa^2Q_a}{2H\kappa^2Q_t}&\epsilon_6&=\frac{\dot Q_t}{2HQ_t}\,
\end{align}
where the auxiliary parameters are given by the following
expressions,
\begin{equation}
\centering
\label{F1}
F=1+2\alpha R=1+24\alpha H^2\, ,
\end{equation}
\begin{equation}
\centering
\label{Q_a1}
Q_a=-8\dot\xi H^2\, ,
\end{equation}
\begin{equation}
\centering
\label{Q_t1}
Q_t=\frac{F}{\kappa^2}-8\dot\xi H\, ,
\end{equation}
\begin{equation}
\centering
\label{E1}
E=\frac{F}{\kappa^2}\left(\omega+\frac{3(\dot F+\kappa^2Q_a)^2}{2\kappa^4\dot\phi^2Q_t}\right)\, ,
\end{equation}
\begin{equation}
\centering
\label{Qe1}
Q_e=-32\dot\xi \dot H\, ,
\end{equation}
The function $Q_e$ does not participate in the slow-roll indices
but is introduced since it participates in subsequent equations.
Concerning $F$, we replaced the Ricci scalar with only Hubble's
parameter squared due to the slow-roll conditions. As a result,
the slow-roll parameters are rewritten as,
\begin{equation}
\centering
\label{index1}
\epsilon_1=\frac{\kappa^2 \omega}{2}\left(\frac{\xi'}{\xi''}\right)^2\, ,
\end{equation}
\begin{equation}
\centering
\label{index2}
\epsilon_2=1-\epsilon_1-\frac{\xi'\xi'''}{\xi''^2}\, ,
\end{equation}
\begin{equation}
\centering
\label{index3}
\epsilon_3=\frac{24\alpha\dot H}{1+24\alpha H^2}\, ,
\end{equation}
\begin{equation}
\centering
\label{index4}
\epsilon_4=\frac{1}{2}\frac{\xi'}{\xi''}\frac{E'}{E}\, ,
\end{equation}
\begin{equation}
\centering
\label{index5}
\epsilon_5=\frac{24\alpha\xi''H^2-4\kappa^2\xi'^2H^2}{\xi''(1+24\alpha H^2)-8\kappa^2\xi'^2H^2}\, ,
\end{equation}
\begin{equation}
\centering \label{index6} \epsilon_6=\frac{24\alpha\xi''\dot
H-4\kappa^2\xi'^2H^2(1-\epsilon_1)}{\xi''(1+24\alpha
H^2)-8\kappa^2\xi'^2H^2}\, .
\end{equation}
Moreover, the auxiliary parameters introduced previously are
rewritten, according to the approximated equations of motion, as
follows,
\begin{equation}
\centering
\label{F2}
F=1+8\alpha\kappa^2V\, ,
\end{equation}
\begin{equation}
\centering
\label{Qa2}
Q_a=-8\frac{\xi'^2}{\xi''}\kappa^2\left(\frac{V}{3}\right)^{\frac{3}{2}}\, ,
\end{equation}
\begin{equation}
\centering
\label{Qt2}
Q_t=\frac{1+8\alpha\kappa^2V}{\kappa^2}-\frac{8}{3}\frac{\xi'^2}{\xi''}\kappa^2V\, ,
\end{equation}
\begin{equation}
\centering \label{E2}
E=\frac{1+8\alpha\kappa^2V}{\kappa^2}\left(\omega+\frac{288\kappa^4\left(\alpha\omega\left(\frac{\xi'}{\xi''}\right)^2\sqrt{\frac{V^3}{3}}+\frac{\xi'^2}{\xi''}\left(\frac{V}{3}\right)^{\frac{3}{2}}\right)}{V\left(\frac{\xi'}{\xi''}\right)^2\left(\frac{1+8\alpha\kappa^2V}{\kappa^2}-\frac{8}{3}\frac{\xi'^2}{\xi''}\kappa^2V\right)}\right)\,
,
\end{equation}

\begin{equation}
\centering \label{Qe2}
Q_e=16\frac{\xi'^4}{\xi''^3}\kappa^5\omega\left(\frac{V}{3}\right)^{\frac{3}{2}}\,
.
\end{equation}
Using the slow-roll indices, we can evaluate the observed indices,
and specifically the spectral index of the primordial scalar
curvature perturbations $n_S$, the tensor-to-scalar ratio $r$ and
the tensor spectral index $n_T$ as,
\begin{align}
\centering
\label{observed}
n_S&=1-2\frac{2\epsilon_1+\epsilon_2-\epsilon_3+\epsilon_4}{1-\epsilon_1}&n_T&=-2\frac{\epsilon_1+\epsilon_6}{1-\epsilon_1}&r&=16\left|\left(\frac{\kappa^2Q_e}{4HF}-\epsilon_1-\epsilon_3\right)\frac{Fc_A^3}{\kappa^2Q_t}\right|\, ,
\end{align}
where $c_A$ denotes the sound wave velocity given by the
expression,
\begin{equation}
\centering \label{cA} c_A^2=1+\frac{\kappa^2Q_e(\dot
F+\kappa^2Q_a)}{2\omega\kappa^4Q_t\dot\phi^2+3(\dot
F+\kappa^2Q_a)^2}\, .
\end{equation}

These are similar equations with the ones used in Ref.
\cite{Odintsov:2020sqy}, but the main difference is that
$\epsilon_3\neq 0$ in the case at hand. This is also the reason
why the exact same models shall be studied, in order to ascertain
the effects of the $R^2$ corrections, and examine how this
correction affects the observational quantities. Since each
function is now written as a function of the scalar field $\phi$,
it shall be used in order to extract information about the
inflationary era. Specifically, by letting
$\epsilon_1\sim\mathcal{O}(1)$, the final value of the scalar
field is derived. Subsequently, by using the $e$-foldings number
defined as $N=\int_{t_i}^{t_f}{Hdt}$ where $t_f-t_i$ signifies the
duration of inflation, and expressing it in terms of the scalar
field $\phi$ as,
\begin{equation}
\centering
\label{efolds}
N=\int_{\phi_i}^{\phi_f}{\frac{\xi''}{\xi'}d\phi}\, ,
\end{equation}
the expression for the scalar field during the first horizon
crossing is produced. Hence, by using it as an input in Eq.
(\ref{observed}), we shall examine whether there exist values for
the free parameters which produce compatible results with the
latest Planck 2018 collaboration \cite{Akrami:2018odb}, which
specifically constrain $n_S$ and $r$ as follows,
\begin{align}
\centering
n_S&=0.9649\pm0.0042&r&<0.064\, ,
\end{align}
with $68\%$C.L and $95\%$C.L respectively. Referring to the tensor
spectral index $n_T$, no specific value is expected since B-Modes
have yet to be observed. In the following, we present the results
for certain model coupling functions which are capable of
simplifying the ratio $\xi'/\xi''$.

An important comment is in order, related to the final form of the
observational indices (\ref{observed}). In our case, a crucial
assumption for the derivation of the observational indices
(\ref{observed}) was the satisfaction of the slow-roll condition
$\epsilon_i\ll 1$, $i=1,3,4,5,6$, that is, for all the
observational indices. A confusion occurs in the literature, since
the authors of Ref. \cite{Hwang:2005hb} derived the expressions
for the observational indices and specifically for the spectral
indices appearing in (\ref{observed}), by using the additional
condition $\dot{\epsilon}_1=0$, or in other texts, this is
modified to the more general $\dot{\epsilon}_1=$const. But these
conditions are superfluous and misleading, and we devote one
article for exactly explaining this delicate issue formally. This
issue was addressed in Ref. \cite{Oikonomou:2020krq} in detail,
where  we demonstrated that the condition $\dot{\epsilon}_1=0$ is
redundant and even can be misleading, and certainly causes
confusion in the literature. As we showed in detail in
\cite{Oikonomou:2020krq}, it is possible to obtain exactly the
same expressions for the spectral indices (tensor and scalar) as
those appearing in Eq. (\ref{observed}), without assuming
$\dot{\epsilon}_1=0$ or even $\dot{\epsilon}_1=$const, by simply
making use of Karamata's theorem for regularly varying functions,
and the slow-roll conditions for the slow-roll indices
$\epsilon_i\ll 1$, $i=1,3,4,5,6$. Thus the inconsistency caused by
the condition $\epsilon_i=$const, is artificial and can be
formally alleviated by using the slow-roll assumptions and
Karamata's theorem.

In fact, the same expression for the spectral index as in Eq.
(\ref{observed}) can be derived without assuming
$\dot{\epsilon}_1=0$, as we showed in Ref.
\cite{Oikonomou:2020krq}.

\section{Testing The Theoretical Framework with the Latest Observational Data}

In this section, we shall ascertain the validity of certain model
functions. Firstly, the Gauss-Bonnet scalar coupling function
shall be defined properly, aiming for a simple ratio $\xi'/\xi''$
as mentioned previously. Afterwards, the scalar potential shall be
derived from Eq. (\ref{motion6}). Ideally, the string corrections
have been neglected from the equations of motion without altering
the results simply because relation (\ref{dotphi}) holds true.
Thus, in the following examples, we shall use a different, more
simplified differential equation for the scalar field as shown
below,
\begin{equation}
\centering
\label{Vdif1}
V'+3\omega H^2\frac{\xi'}{\xi''}\simeq0\, ,
\end{equation}
which is valid only if $24\dot\xi H^4$ is indeed negligible
compared to the rest terms. This new form is inspired from
neglecting of $24\dot\xi H^3$ and $16\dot\xi H\dot H$ from the
first and second equation of motion in (\ref{motion1}) and
(\ref{motion2}) respectively, and it leads to a much more elegant
functional form of the scalar potential as we shall demonstrate in
the following examples. Furthermore, the opposite case is also
interesting, where some of the string corrections in the
differential equation are kept while the kinetic term is omitted,
meaning that,
\begin{equation}
\centering
\label{Vdif2}
V'+24\xi' H^4\simeq0\, ,
\end{equation}
due to the fact that $H$ is proportional to the scalar potential
$V$ and also $\kappa\xi'/\xi''\ll1$, hence once obtains an
ordinary differential equation, however this case will not be
studied in the present paper. In the following we demonstrate
certain viable models.

\subsection{Error Function Coupling Under The Slow-Roll Assumption}

We begin our examples by designating the Gauss-Bonnet coupling as
follows,
\begin{equation}
\centering
\label{xiA}
\xi(\phi)=\frac{2\lambda_1}{\sqrt{\pi}}\int_{0}^{\gamma_1\kappa\phi}{e^{-x^2}dx}\, ,
\end{equation}
where $\lambda_1$ and $\gamma_1$ are dimensionless constants to be
specified later, whereas $x$ is an auxiliary integration variable.
In Ref. \cite{Odintsov:2020sqy}, the complete differential
equation (\ref{motion6}) was used and the produced scalar
potential was at the least lengthy. Since Hubble's parameter and
the differential equation for the potential have not been altered
in the $R^2$ case, as it was demonstrated in the previous section,
the same potential is sure to be produced in this case as well.
Hence, we shall use Eq. (\ref{Vdif1}) in order to examine whether
the string corrections are as negligible as we stated they were.
Using Eq. (\ref{Vdif1}), the scalar potential reads,
\begin{equation}
\centering
\label{V1}
V(\phi)=V_1(\kappa\phi)^{\frac{\omega}{2\gamma_1^2}}\, ,
\end{equation}
where $V_1$ is an arbitrary integration constant with mass
dimensions $[m]^4$ and signifies the amplitude of the potential.
Hence, by omitting the string corrections from Eq.
(\ref{motion6}), one obtains a simple power-law form for the
scalar potential with the exponent not necessarily being an
integer. Furthermore, $F$ might not be directly present in the
equations of motion, however, it alters the dynamics of the model
as it participates in the observational indices, hence a
difference between the pure Einstein-Gauss-Bonnet case and the
$R^2$ approach shall be demonstrated. Concerning the slow-roll
indices, these are,
\begin{equation}
\centering
\label{index1A}
\epsilon_1=\frac{\omega}{2(2\gamma_1^2\kappa\phi)^2}\, ,
\end{equation}
\begin{equation}
\centering
\label{index2A}
\epsilon_2=\frac{ (2 \gamma_1)^2-\omega}{2(2 \gamma_1^2 \kappa  \phi) ^2}\, ,
\end{equation}
\begin{equation}
\centering
\label{index3A}
\epsilon_3=-\frac{\alpha  \omega \kappa^2 V(\phi )}{(\gamma_1^2 \kappa\phi) ^2 \left(8 \alpha  \kappa ^2 V(\phi )+1\right)}\, ,
\end{equation}
\begin{equation}
\centering
\label{index5A}
\epsilon_5=\frac{\kappa^2V(\phi ) \left(4\gamma_1^3 \kappa ^2\lambda_1 \kappa\phi -3 \sqrt{\pi } \alpha  \omega  e^{(\gamma _1 \kappa  \phi)^2}\right)}{\gamma_1^3\kappa \phi  \left(3 \sqrt{\pi }\gamma_1 \kappa\phi  e^{(\gamma_1\kappa \phi )^2}+8 \kappa ^2 V(\phi ) \left(3 \sqrt{\pi } \alpha \gamma_1 \kappa\phi  e^{(\gamma_1 \kappa \phi)^2}+\kappa^2 \lambda_1\right)\right)}\, ,
\end{equation}
\begin{equation}
\centering
\label{index6A}
\epsilon_6=\frac{2  \lambda_1\kappa^4 V(\phi ) \left(2 (\gamma_1\kappa\phi) ^2+1\right)-2\kappa \phi\kappa  V'(\phi ) \left(3 \sqrt{\pi } \alpha  \gamma_1\kappa \phi  e^{(\gamma_1 \kappa  \phi) ^2}+\kappa^2  \lambda_1\right)}{(\gamma_1\kappa\phi) ^2 \left(3 \sqrt{\pi } \gamma_1 \kappa\phi  e^{(\gamma_1\kappa  \phi) ^2}+8 \kappa ^2 V(\phi ) \left(3 \sqrt{\pi } \alpha  \gamma_1 \kappa\phi  e^{(\gamma_1 \kappa  \phi) ^2}+\kappa^2  \lambda_1\right)\right)}\, ,
\end{equation}
It is clear that the first three slow-roll indices have simple
functional forms, due to the simplified ratio $\xi'/\xi''$,
however this is not the case for $\epsilon_4$, since the produced
form is quite lengthy, it was deemed suitable to neglect such
index but it still participates in the scalar spectral index.
Utilizing the first slow-roll index and the $e$-foldings number,
meaning equations (\ref{index1A}) and (\ref{efolds}), the initial
and final value of the scalar field read,
\begin{equation}
\centering
\label{phiiA}
\phi_i=\pm\frac{\sqrt{N+(\gamma_1\kappa\phi_f)^2}}{\gamma_1\kappa}\, ,
\end{equation}
\begin{equation}
\centering
\label{phifA}
\phi_f=\pm\sqrt{\frac{\omega}{2}}\frac{1}{2\gamma_1^2\kappa}\, ,
\end{equation}
In the following, we shall limit our work only in the positive
values of the scalar field. Assigning the following values for the
free parameters of the model in Planck Units, meaning
$\kappa^2=1$, ($\omega$, $\lambda_1$, $N$, $\gamma_1$, $V_1$,
$\alpha$)=(1, 1, 60, 0.5, 1, $10^{-3}$), then the resulting
observational indices are equal to $n_S=0.966841$, $r=0.045034$
and $n_T=-0.005677$ which are obviously acceptable values. In
addition, since $c_A=1$, the model is free of ghost instabilities.
Lastly, we mention that the numerical values of the slow-roll
indices, which are indicative of the validity of the slow-roll
conditions, are $\epsilon_1=0.0082644$,
$\epsilon_2=3.3\cdot10^{-16}$, $\epsilon_3=-0.00545$,
$\epsilon_4=-0.005534$ and $\epsilon_5,\epsilon_6$ are both equal
to $\epsilon_3$. This in turn implies that the slow-roll
conditions hold true. Also in Fig. 1 we plot the
spectral index of primordial curvature perturbations $n_S$ (left)
and the tensor-to-scalar ratio $r$ (right) depending on parameters
$\gamma_1$ and $V_1$ ranging from [0.5,2] and [0.1,2]
respectively.
\begin{figure}[h!]
\centering
\includegraphics[width=17pc]{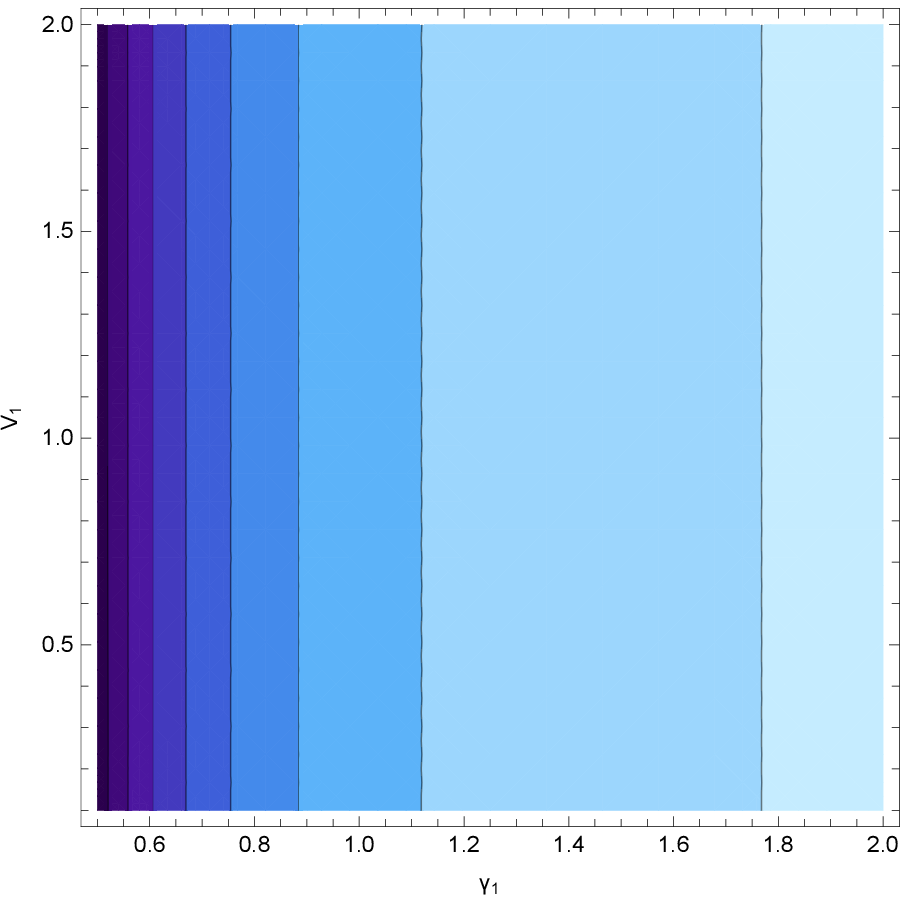}
\includegraphics[width=3pc]{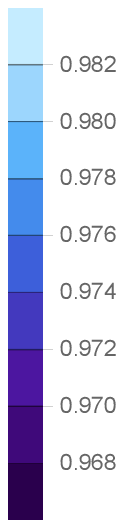}
\includegraphics[width=17pc]{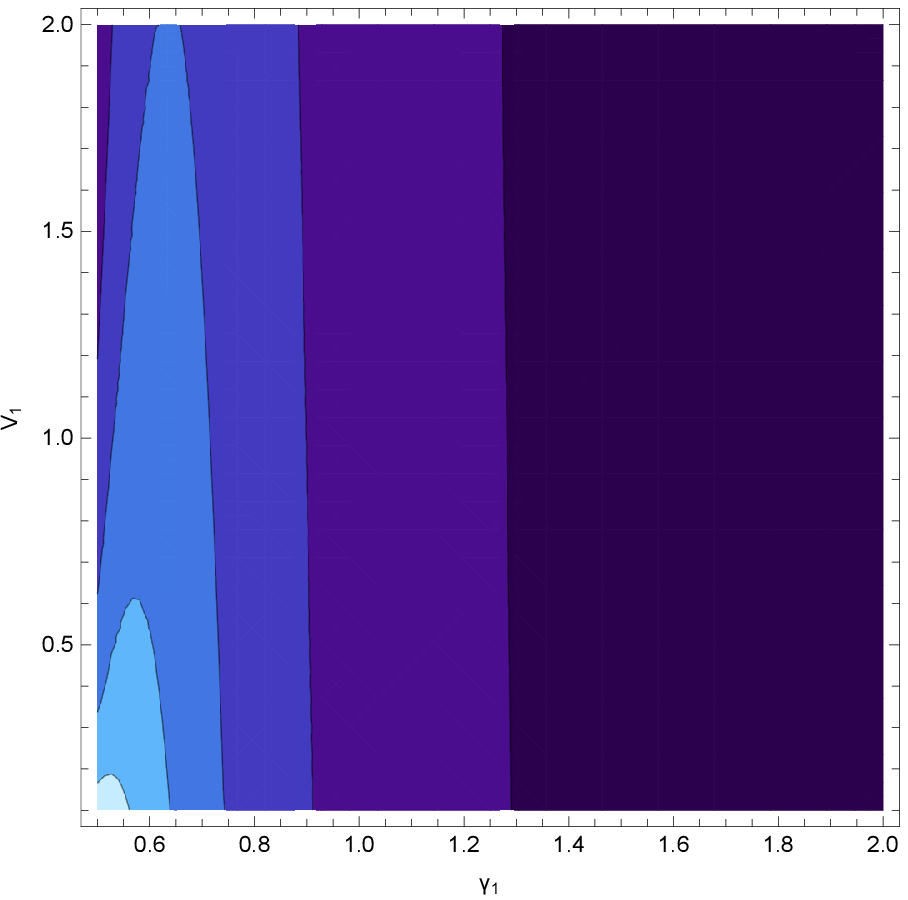}
\includegraphics[width=2.5pc]{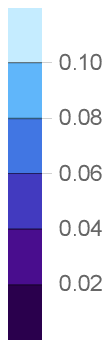}
\caption{Spectral index of primordial curvature perturbations
$n_S$ (left) and tensor-to-scalar ratio $r$ (right) depending on
parameters $\gamma_1$ and $V_1$ ranging from [0.5,2] and [0.1,2]
respectively.}
\end{figure}
It is interesting to make comparisons between the $R^2$-corrected
Einstein-Gauss-Bonnet gravity studied in this paper with the pure
Einstein studied in Ref \cite{Odintsov:2020sqy}. Firstly, the
scalar potential in this case is extremely simple, as it turns out
that $V(\phi)\sim(\kappa\phi)^2$ for the previous designation.
Hence, the exponent is indeed integer. The major difference is
that the previous designation is not fine tunined, which was not
the case in Ref. \cite{Odintsov:2020sqy}. In that case, the
integration constant $c$ of Ref. \cite{Odintsov:2020sqy}, which in
this case can be though of as the inverse of the potential
amplitude, was of order $\mathcal{O}(10^{-25})$. In the present
paper, the only parameter with a small value happens to be
$\alpha$. This particular designation is somewhat necessary in
order to not only extract compatible results, but also satisfy the
approximations which were imposed previously. The spectral index
between the case studied in \cite{Odintsov:2020sqy} and the
$R^2$-corrected case remains the same and as a matter of fact, the
main difference is the tensor-to-scalar ratio, which in the case
at hand is greater than the case studied in Ref.
\cite{Odintsov:2020sqy}.

For parameter $\alpha$, we need to note that in order for Eq.
(\ref{motion5}) to be valid, $\alpha$ must be approximately of
order $\mathcal{O}(10^{-3})$ and smaller, otherwise $\ddot F$
becomes equal to, or greater than, the kinetic term
$\omega\dot\phi^2$ in Eq. (\ref{motion2}).

It is also worth mentioning that changing parameter $\alpha$ to
$\alpha=0.019$ produces observational indices which both respect
the approximate expressions of the $R^2$ gravity, however the
approximations in the equation of motion are violated,
specifically $144\alpha H\dot H$ and $V$ are of the same order
while $\ddot F$ and $\omega\dot\phi^2$ are differing only by one
order of magnitude, hence we avoided to give these values to the
parameter $\alpha$.

Furthermore, it is worthy to discuss the correspondence of the
system under a change in a free parameter. For $\gamma_1$, an
increase leads to a subsequent increase in $n_S$ while
simultaneously $r$ decreases. In addition, the amplitude of the
scalar potential manages to increase $n_S$ and decrease $r$,
faster than $\gamma_1$, once $V_1$ increases. The opposite
obviously happens when either $\gamma_1$ or $V_1$ decreases.
Keeping the current values for the free parameters and changing
only $\alpha$, we stated that a decrease manages to produce
incompatible results while an increase makes the approximations in
the equations of motion invalid. Although this is true, we mention
that changing both $\alpha$ and $V_1$, for instance
$\alpha=10^{-4}$ and $V_1=10$ produces the exact same
observational indices, however, the order of magnitude of each
term in the equations of motion experiences the same change, hence
their ratio's remain the same.

Finally, we mention that all the approximations imposed in the
equations of motion, meaning the slow-roll conditions and the
string corrections are justifiable, for the values of the free
parameters we used in order to obtain viability of the model with
the observational data. Firstly, as it was hinted from the
numerical values of the slow-roll indices, we mention that the
slow-roll approximations hold since $\dot
H\sim\mathcal{O}(10^{-1})$ while $H^2\sim\mathcal{O}(10)$,
$\frac{1}{2}\omega\dot\phi^2\sim\mathcal{O}(10^{-1})$ while
$V\sim\mathcal{O}(100)$ and finally,
$\ddot\phi\sim\mathcal{O}(10^{-17})$ whereas
$H\dot\phi\sim\mathcal{O}(10)$. Furthermore, the following terms
are indeed negligible since $24\dot\xi
H^3\sim\mathcal{O}(10^{-23})$, $16\dot\xi H\dot
H\sim\mathcal{O}(10^{-25})$ and
$24\xi'H^4\sim\mathcal{O}(10^{-22})$ while
$V'\sim\mathcal{O}(10)$. Thus, it was reasonable to assume that
(\ref{Vdif1}) is the right choice for producing the scalar
potential. Finally, since $\alpha=10^{-3}$, as stated before, we
mention that $144\alpha H^2\dot H\sim\mathcal{O}(1)$, $\ddot
F\sim\mathcal{O}(10^{-2})$ thus neglecting such terms relative to
$V$ and $\omega\dot\phi^2$ is justifiable.

\subsection{Advanced Exponential As a Coupling Function Under The Slow-Roll Assumption}

Let us now consider another model, so in this case we assume that,
\begin{equation}
\centering
\label{xiB}
\xi(\phi)=\kappa\lambda_2\int_{0}^{\kappa\phi}{e^{-\gamma_2 x^2}dx}\, ,
\end{equation}
where now $\lambda_2$ and $\gamma_2$ are now the dimensionless
parameters of the model. This choice also simplifies the ratio
$\xi'/\xi''$ hence the selection. Using Eq. (\ref{Vdif1}), the
resulting scalar potential reads,
\begin{equation}
\centering
\label{VB}
V(\phi)=V_2 e^{-\frac{(\kappa\phi )^{2-m} \omega }{\gamma_2 (2-m) m}}\, ,
\end{equation}
where as before, $V_2$ is the integration constant with mass
dimensions $[m]^4$. In this case, the scalar potential has an
exponential form and not a power-law form as before, however in
this case as well the produced potential is very simple
functionally. As a result, the slow-roll indices take the
following expressions,
\begin{equation}
\centering
\label{index1B}
\epsilon_1=\frac{\omega(\kappa\phi)^{2-2m}}{2(m\gamma_2)^2}\, ,
\end{equation}
\begin{equation}
\centering
\label{index2B}
\epsilon_2=-\frac{\omega(\kappa\phi)^2+2m\gamma_2(m-1)(\kappa\phi)^m}{2m^2\gamma_2^2(\kappa\phi)^{2m}}\, ,
\end{equation}
\begin{equation}
\centering
\label{index3B}
\epsilon_3=-\frac{4\alpha\omega(\kappa\phi)^{2-m}\kappa^2V}{(m\gamma_2)^2(1+8\alpha\kappa^2V)}\, ,
\end{equation}
\begin{equation}
\centering \label{index5B} \epsilon_5=-\frac{4 \kappa ^4 V(\phi )
(\kappa  \phi )^{-m} \left(3 \alpha  \phi^2  \omega +\gamma_2
\kappa\phi  \lambda_2 m (\kappa  \phi )^m e^{\gamma_2 (\kappa \phi
)^m}\right)}{\gamma_2 m \left(3 \gamma_2 m (\kappa  \phi )^m-8
\kappa ^2 V(\phi ) \left(\kappa ^3 \lambda_2 \phi e^{\gamma_2
(\kappa  \phi )^m}-3 \alpha  \gamma_2 m (\kappa  \phi
)^m\right)\right)}\, .
\end{equation}
Once again, the first three slow-roll indices have simple and
elegant forms whereas the rest are intricate. From the first
slow-roll index however, one obtains the following values for the
scalar field,
\begin{equation}
\centering
\label{phiiB}
\phi_i=\frac{1}{\kappa}\left((\kappa\phi_f)^m-\frac{N}{\gamma_2}\right)^{\frac{1}{m}}\, ,
\end{equation}
\begin{equation}
\centering \label{phifB}
\phi_f=\frac{1}{\kappa}\left(\frac{2m^2\gamma_2^2}{\omega}\right)^{\frac{1}{2-2m}}\,
.
\end{equation}

By giving the following values for the free parameters in reduced
Planck units $(\omega$, $\lambda_2$, $N$, $\gamma_2$, $V_2$, $m$,
$\alpha$)=(1, -1, 60, -0.5, 1, -6, $10^{-3}$), then the
observational indices obtain the values $n_S=0.961203$,
$n_T=-1.5\cdot10^{-6}$ and $r=1.23\cdot10^{-5}$ so once again the
results are compatible. Additionally, $c_A=1$ thus the model is
free of ghosts and finally, $\epsilon_1=7\cdot10^{-7}$,
$\epsilon_2=0.0165$ and indices $\epsilon_3$ through $\epsilon_6$
are equal to $-8\cdot10^{-11}$. Hence, the slow-roll conditions
indeed apply, in the case of the second index however they are
marginally applicable. In Fig. 2 we plot the spectral
index of primordial curvature perturbations $n_S$ (left) and the
tensor-to-scalar ratio $r$ (right) depending on parameters
$\gamma_2$ and $m$ ranging from [-2,-0.5] and [-10,-4]
respectively, and we can see that the viability of the model can
be ensured for a wide range of the parameters used.
\begin{figure}[h!]
\centering
\includegraphics[width=17pc]{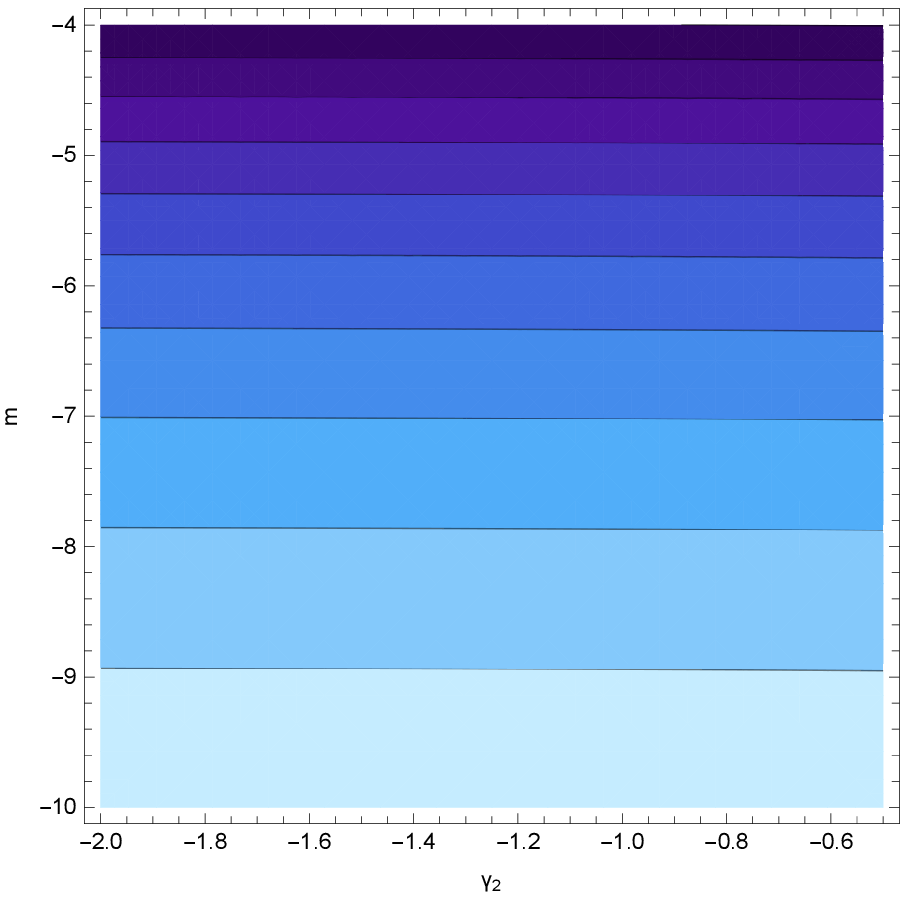}
\includegraphics[width=3pc]{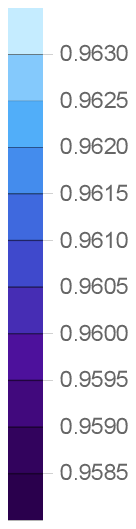}
\includegraphics[width=17pc]{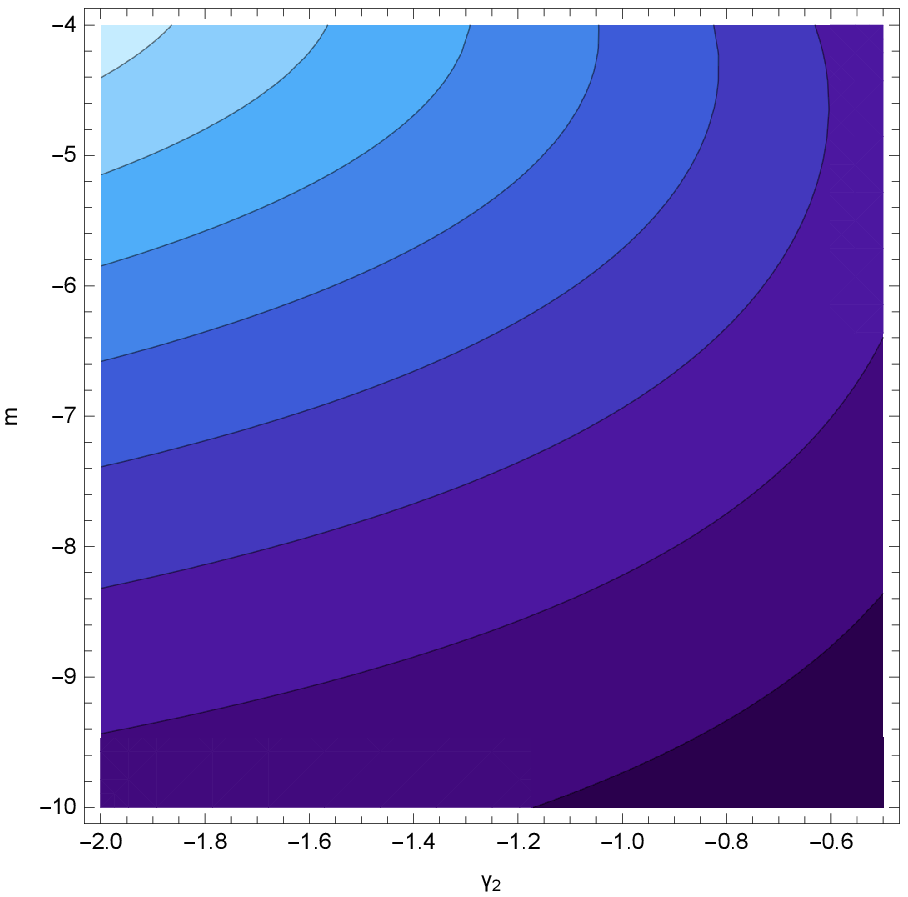}
\includegraphics[width=3.5pc]{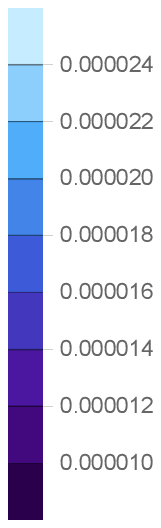}
\caption{Spectral index of primordial curvature perturbations
$n_S$ (left) and tensor-to-scalar ratio $r$ (right) depending on
parameters $\gamma_2$ and $m$ ranging from [-2,-0.5] and [-10,-4]
respectively.}
\end{figure}
We need to note that the approximations we made are valid, but
parameter $\alpha$ must be of order $\mathcal{O}(10^{-3})$ or
smaller in order for the approximations to be valid. When it comes
to the rest parameters, the situation is similar to the previous
model. This was also the reason this coupling function was chosen,
as it shares some common characteristics with the error function.

Finally, we discuss the validity of the approximations imposed.
When it comes to the slow-roll approximations, we mention that
$\dot H\sim\mathcal{O}(10^{-7})$ while
$H^2\sim\mathcal{O}(10^{-1})$,
$\frac{1}{2}\omega\dot\phi^2\sim\mathcal{O}(10^{-7})$ while
$V\sim\mathcal{O}(1)$ and lastly
$\ddot\phi\sim\mathcal{O}(10^{-6})$ whereas
$H\dot\phi\sim\mathcal{O}(10^{-4})$ so indeed the slow-roll
approximations hold when the free parameters of the theory obtain
the previous values. Concerning the additional approximations,
$24\dot\xi H^3\sim\mathcal{O}(10^{-29})$, $16\dot\xi H\dot
H\sim\mathcal{O}(10^{-35})$ and
$24\xi'H^4\sim\mathcal{O}(10^{-26})$ compared to
$V'\sim\mathcal{O}(10^{-3})$. Moreover, concerning the derivatives
of $F$, we have $144\alpha H^2\dot H\sim\mathcal{O}(10^{-8})$ and
$\ddot F\sim\mathcal{O}(10^{-10})$ thus neglecting such terms
compared to the kinetic term is justifiable.

\subsection{Power-Law Coupling Under the Constant-Roll Assumption}

In this final example, we shall assume that the Gauss-Bonnet
coupling function is given by the following expression,
\begin{equation}
\centering
\label{xiC}
\xi(\phi)=\lambda_3(\kappa\phi)^n\, ,
\end{equation}
Here, we shall also assume that the scalar field obeys the
constant-roll condition $\ddot\phi=\beta H\dot\phi$ where $\beta$
is the constant-roll parameter to be specified later. As a result,
the condition for the time derivative of the scalar field, derived
from Eq. (\ref{cT}) reads,
\begin{equation}
\centering
\label{dotphi2}
\dot\phi=H(1-\beta)\frac{\xi'}{\xi''}\, ,
\end{equation}
Substituting $\beta=0$ restores the previous equations. Also, it
is worth stating that exponent $n$ cannot obtain the value $n=1$,
implying that the Gauss-Bonnet coupling cannot be a linear
function of $\phi$. This is because the constant-roll parameter
becomes equal to $\beta=1$ hence the slow-roll conditions cannot
be implemented. Consequently, the equations of motion
(\ref{motion4}) through (\ref{motion6}) are written as,
\begin{equation}
\centering
\label{motion4C}
H^2=\frac{\kappa^2V}{3}\, ,
\end{equation}
\begin{equation}
\centering
\label{motion5C}
\dot H=-H^2\frac{(1-\beta)^2\kappa^2\omega}{2}\left(\frac{\xi'}{\xi''}\right)^2\, ,
\end{equation}
\begin{equation}
\centering
\label{motion6C}
V'+(3+\beta)(1-\beta)\omega H^2\frac{\xi'}{\xi''}=0\, ,
\end{equation}
Thus, only the last two equations were affected by the
constant-roll condition directly, however since the scalar
potential is derivable from Eq. (\ref{motion6C}), then Eq.
(\ref{motion4C}) shall also depend on $\beta$. Furthermore, the
slow-roll indices in this framework, for an unspecified
Gauss-Bonnet coupling function, are given by the following
expressions,
\begin{equation}
\centering
\label{index1C}
\epsilon_1=\frac{(1-\beta)^2\kappa^2\omega}{2}\left(\frac{\xi'}{\xi''}\right)^2\, ,
\end{equation}
\begin{equation}
\centering
\label{index2C}
\epsilon_2=\beta\, ,
\end{equation}
\begin{equation}
\centering
\label{index3C}
\epsilon_3=\frac{24\alpha\dot H}{1+24\alpha H^2}\, ,
\end{equation}
\begin{equation}
\centering
\label{index4C}
\epsilon_4=\frac{1-\beta}{2}\frac{\xi'}{\xi''}\frac{E'}{E}\, ,
\end{equation}
\begin{equation}
\centering
\label{index5C}
\epsilon_5=\frac{24\alpha\xi''H^2-4(1-\beta)\kappa^2\xi'^2H^2}{\xi''(1+24\alpha H^2)-8(1-\beta)\kappa^2\xi'^2H^2}\, ,
\end{equation}
\begin{equation}
\centering
\label{index6C}
\epsilon_6=\frac{24\alpha\xi''H^2-4(1-\beta)\kappa^2\xi'^2H^2(1-\epsilon_1)}{\xi''(1+24\alpha H^2)-8(1-\beta)\kappa^2\xi'^2H^2}\, ,
\end{equation}
We shall refrain from rewriting the auxiliary parameters $Q_i$ and
$E$ since they differ only by a factor of $1-\beta$, but we
mention that the $e$-foldings number is given by the expression,
\begin{equation}
\centering
N=\frac{1}{1-\beta}\int_{\phi_i}^{\phi_f}{\frac{\xi''}{\xi'}d\phi}\,
.
\end{equation}

Let us proceed with the model at hand. From equations (\ref{xiC})
and (\ref{motion6C}), one obtains the following scalar potential,
\begin{equation}
\centering
\label{VC}
V(\phi)=V_3e^{\frac{\omega(\beta^2+2\beta-3)(\kappa\phi)^2}{6(n-1)}}\,
\end{equation}
where $V_3$ is the integration constant with mass dimensions
$[m]^{4}$ for consistency. The resulting potential is a
exponential and has a simple functional form. Consequently, the
slow-roll indices for this specific pair of scalar functions are
written as follows,
\begin{equation}
\centering
\label{index1C1}
\epsilon_1=\frac{\omega}{2}\left(\frac{(1-\beta)\kappa\phi}{n-1}\right)^2\, ,
\end{equation}
\begin{equation}
\centering
\label{index2C1}
\epsilon_2=\beta\, ,
\end{equation}
\begin{equation}
\centering
\label{index3C1}
\epsilon_3=-\frac{4(1-\beta)^2\alpha\phi^2\omega\kappa^4V}{(n-1)^2(1+8\alpha\kappa^2V)}\, ,
\end{equation}
\begin{equation}
\centering \label{index5C1} \epsilon_5=-\frac{4 (1-\beta ) \kappa
^4 V(\phi ) \left(3 \alpha (1-\beta ) \phi ^2 \omega
+n(n-1)\lambda_3(\kappa  \phi )^n\right)}{(n-1) \left(8 \kappa ^2
V(\phi ) \left(3 \alpha (n-1)-(1-\beta ) \kappa ^2 \lambda_3 n
(\kappa  \phi )^n\right)+3 (n-1)\right)}\, .
\end{equation}
From Eq. (\ref{index1C1}) and Eq. (\ref{efolds}), the expressions
for the initial and final value of the scalar field are derived,
which in this case read,
\begin{equation}
\centering
\label{phiiC}
\phi_i=\phi_f e^{-\frac{N(1-\beta)}{n-1}}\, ,
\end{equation}
\begin{equation}
\centering
\label{phifC}
\phi_f=\pm\frac{1}{\kappa}\sqrt{\frac{2}{\omega}}\left|\frac{n-1}{1-\beta}\right|\, ,
\end{equation}
In the following, we shall limit our work only to the positive
value of the scalar field. Assigning the following values to the
free parameters, always in reduced Planck units, ($\omega$,
$\lambda_3$, $N$, $V_3$, $\alpha$, $n$, $\beta$)=(1, 1, 60, 1,
0.001, 15, 0.0165) then the observational indices take the values
$n_S=0.96612$, $r=0.00346392$ and $n_T=-0.000433029$ which are
obviously compatible with the latest Planck data
\cite{Akrami:2018odb}. Concerning the scalar field, we mention
that $\phi_i=0.297384$ and $\phi_f=20.1312$ hence an increasing
with time field is present. Also, as expected $c_A=1$ and lastly,
$\epsilon_1=0.000218$, $\epsilon_3=-1.7\cdot10^{-6}$ and the rest
indices are approximately equal to $\epsilon_3$ hence the
slow-roll conditions are valid assumptions. In Fig. 3 we
plot the spectral index of primordial curvature perturbations
$n_S$ (left) and tensor-to-scalar ratio $r$ (right) depending on
exponent $n$ and constant-roll parameter $\beta$, ranging from
[10,17] and [0.015, 0.02] respectively. As it can be inferred,
there exist multiple pairs which lead to viable results.
\begin{figure}[h!]
\centering
\includegraphics[width=17pc]{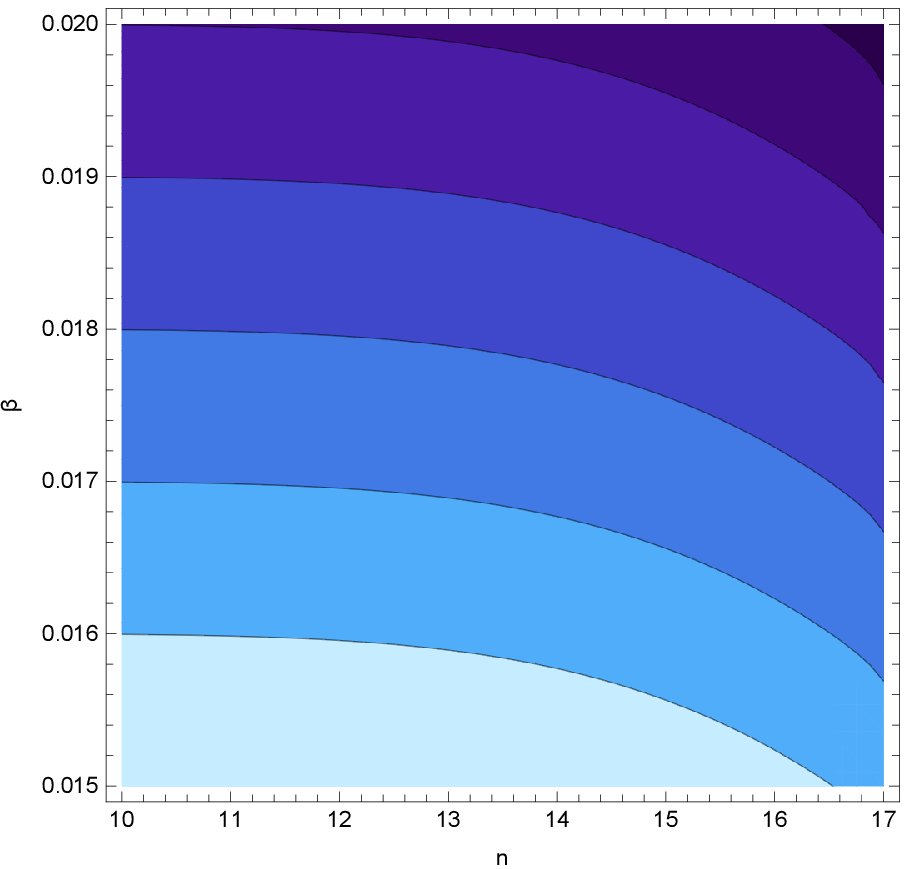}
\includegraphics[width=2.8pc]{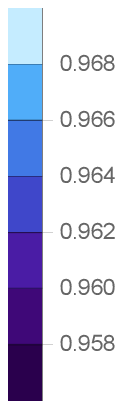}
\includegraphics[width=17pc]{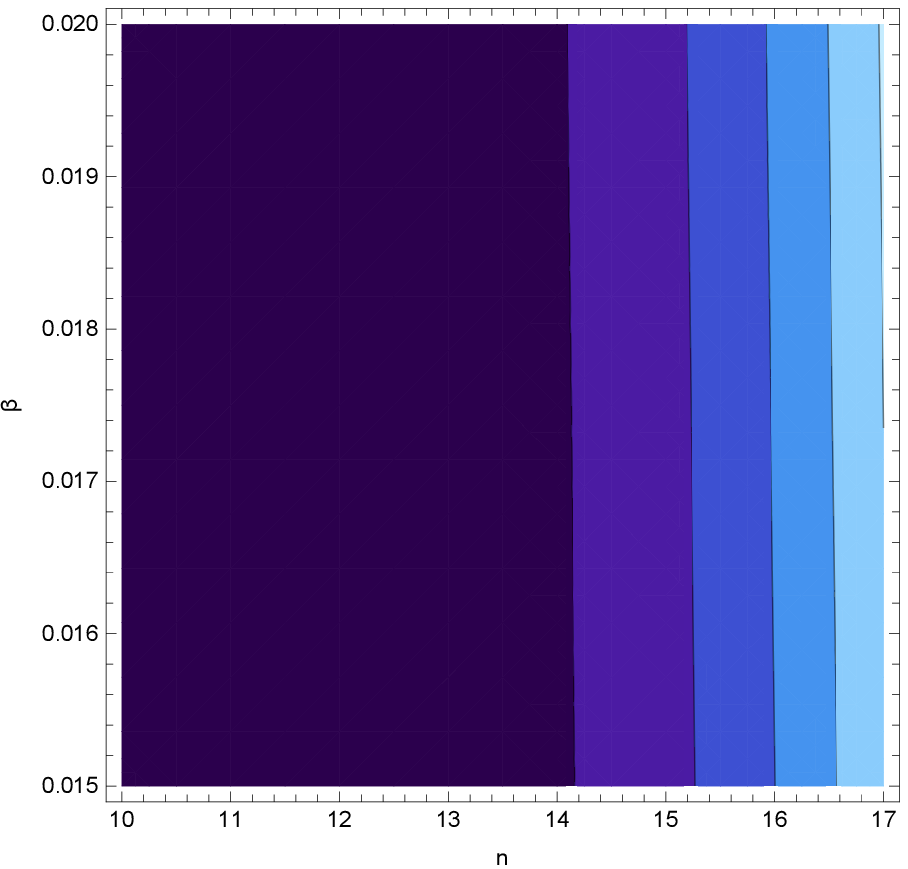}
\includegraphics[width=3pc]{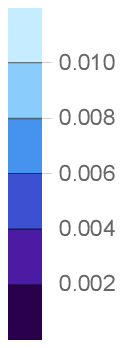}
\caption{Spectral index of primordial curvature perturbations
$n_S$ (left) and tensor-to-scalar ratio $r$ (right) depending on
exponent $n$ and constant-roll parameter $\beta$, ranging from
[10,17] and [0.015, 0.02] respectively. As it can be inferred,
there exist multiple pairs which lead to viable results.}
\end{figure}
In this simple model once again $\alpha$ must be equal to or
smaller that $10^{-3}$ in order for the approximated equations of
motion to be valid. Also, the constant-roll parameter $\beta$
affects the spectral index of primordial curvature perturbations
greatly while the tensor-to-scalar ratio mildly. Specifically, a
decrease in $\beta$ leads to an increase in both $n_S$ and $r$. In
contrast, the exponent $n$ affects the tensor-to-scalar ratio more
than the scalar spectral index. Decreasing $n$ leads to a decrease
in $r$ while it increases $n_S$. Also, $\alpha$ affects the
tensor-to-scalar ratio, but it is not a significant change. For
instance, letting $\alpha=10^{-5}$, a decrease in order to be
consistent with the approximations in the equations of motion,
leads to a numerical change in the fifth decimal of $r$. Moreover,
$V_3$ affects once again only the tensor-to-scalar ratio.
Assigning $V_3=10$ leads to $r=0.003234$ while $V_3=0.1$ results
in $r=0.003489$. The parameter $\lambda_3$ does not affect the
observational indices, but is used only to decrease the order of
magnitude of the string corrections.

Finally, we discuss the validity of the approximations which were
made in the equations of motion. Referring to the slow-roll
conditions, we mention that $\dot H\sim\mathcal{O}(10^{-5})$
whereas $H^2\sim\mathcal{O}(10^{-1})$ and
$\frac{1}{2}\omega\dot\phi^2\sim\mathcal{O}(10^{-5})$ while
$V\sim\mathcal{O}(1)$. In addition, the following terms are indeed
negligible since $24\dot\xi H^3\sim\mathcal{O}(10^{-8})$,
$16\dot\xi H\dot H\sim\mathcal{O}(10^{-12})$ and
$24\xi'H^4\sim\mathcal{O}(10^{-6})$ in comparison to
$V'\sim\mathcal{O}(10^{-2})$. Their order of magnitude can also
further be decreased by altering $\lambda_3$, and specifically by
decreasing $\lambda_3$. Lastly, we mention that $144\alpha H^2\dot
H\sim\mathcal{O}(10^{-6})$ while $\ddot F\sim\mathcal{O}(10^{-7})$
thus it is reasonable to neglect such terms. Moreover, a decrease
in $\alpha$ leads to a subsequent decrease in these terms as well.

\section{The Ghost Instabilities Issue: Are There any Ghost Modes}

Let us now discuss an important issue for the $R^2$- and
string-corrected canonical scalar theory we studied in this paper,
and specifically the ghost issue. So the question is , are there
any ghost propagating modes? At the cosmological level, this issue
is easily answered by examining the values that the sound wave
velocity $c_A$ appearing in Eq. (\ref{cA}) can take. Specifically,
tachyonic propagating modes can occur if the sound speed $c_A$
takes negative values or values $c_A>1$. Let us recall how the
sound speed enters the cosmological perturbations evolution
equations, and how it determines whether ghost modes occur or not.
The cosmological perturbations are obtained by actually deforming
the flat FRW background $g_{\alpha \beta}^{(3)}$ as follows
\cite{Hwang:2005hb},
\begin{equation}\label{perturbationscosmo}
ds^2=-a^2(1+\alpha)d\eta^2-2\alpha^2\beta_{,\alpha}d \eta
dx^{\alpha}+a^2\left(g_{\alpha \beta}^{(3)}+2\varphi g_{\alpha
\beta}^{(3)}+2\gamma_{,\alpha | \beta}+2C_{\alpha \beta}
\right)dx^{\alpha}dx^{\beta}\, ,
\end{equation}
where the FRW background $g_{\alpha \beta}^{(3)}$ is,
\begin{equation}\label{flatfrwanalytic}
g_{\alpha
\beta}^{(3)}dx^{\alpha}dx^{\beta}=dr^2+r^2(d\theta^2+\sin^2 \theta
d\phi^2)\, ,
\end{equation}
and also $\eta$ and $a$ in Eq. (\ref{perturbationscosmo}) stand
for the conformal time and the scale factor respectively. Also
$\alpha$, $\beta$, $\gamma$ and $\varphi$ in Eq.
(\ref{perturbationscosmo}) are scalar type order variables of the
perturbations, while $C_{\alpha \beta}$ is the tensor perturbation
is trace-free and transverse. The evolution of the scalar
perturbation variable $\Phi=\varphi_{\delta \phi}$, which is
related to the scalar gauge invariant variable $\delta
\phi_{\varphi}=-\frac{\dot{\phi}}{H}\varphi_{\delta \phi}$, for
the case of the $R^2$- and string-corrected canonical scalar
theory, is governed by the following differential equation,
\begin{equation}\label{bigdiffeeqannew}
\frac{\left(H+\frac{\dot{F}+Q_a}{2F+Q_b}\right)^2}{a^3\left(\omega\dot{\phi}^2+3\frac{(\dot{F}+Q_a)^2}{2F+Q_b}+Q_c
\right)}\frac{d}{d t} \left
(\frac{a^3\left(\omega\dot{\phi}^2+3\frac{(\dot{F}+Q_a)^2}{2F+Q_b}+Q_c\right)}{\left(H+\frac{\dot{F}+Q_a}{2F+Q_b}\right)^2}
\dot{\Phi}\right )=c_A^2\frac{\Delta}{a^2}\Phi\, ,
\end{equation}
where $\Delta$ is the Laplacian operator corresponding to the
spatial section of the FRW metric, and $c_A$ is the sound wave
speed defined in Eq. (\ref{cA}). As it is obvious from the above
differential equation, ghost propagating modes can occur if the
sound wave speed $c_A$, can develop negative values or values
$c_A>1$. We performed a full numerical analysis for the values of
$c_A$ for all the models we studied in this paper, and for the
values of the free parameters that yield compatibility of the
theory with the observational constraints for inflation coming
from the Planck data. Our analysis showed that the sound wave
speed is remarkably equal to unity for all the models, as we
already mentioned in the text too. Let us here examine one case,
just for illustrative purposes. We shall consider the advanced
exponential model of the next to previous subsection, and let us
see how the sound speed $c_A$ behaves as we vary for example the
parameter $\gamma_2$, so in Fig. 4 we plot the behavior
of the sound wave speed $c_A$ as a function of the parameter
$\gamma_2$, for $(\omega$, $\lambda_2$, $N$, $\gamma_2$, $V_2$,
$m$, $\alpha$)=(1, -1, 60, 1, -6, $10^{-3}$) and , and recall that
these values of the free parameters the inflationary phenomenology
of this specific model is rendered viable. As shown in Fig.
4, the sound wave speed is equal to unity, thus no ghost
instabilities occur for this model. The same results hold true for
the other models we studied in this paper, but we do not present
these here for brevity.
\begin{figure}[h!]
\centering \label{extra}
\includegraphics[width=17pc]{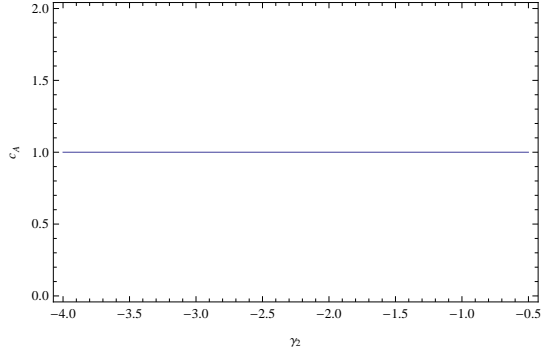}
\caption{The sound wave speed $c_A$ as a function of $\gamma_2$,
for $(\omega$, $\lambda_2$, $N$, $\gamma_2$, $V_2$, $m$,
$\alpha$)=(1, -1, 60, 1, -6, $10^{-3}$) and for
$\gamma_2=[-2,-0.5]$.}
\end{figure}

We should note that the analytic study of the wave speed is not
feasible for the model under study, due to the complexity of the
resulting functional form of the wave speed $c_A$. Let us see why,
the wave speed for the present model has the form,
\begin{align}\label{wavespeedanalyticalexpr}
&c_A= \\ \notag &\frac{\sqrt{\frac{32 V(\phi )^3  \xi
'(\phi )^3 V'(\phi )^2 \left(27 \alpha ^2-2 \xi
'(\phi )^2\right)+192 \alpha  V(\phi )^2 \xi '(\phi )^4
V'(\phi )^3+24 V(\phi )^3 \xi '(\phi )^4 \xi '(\phi)
 (1-24 \alpha  V(\phi )) V'(\phi )+72\alpha V(\phi )^4 \xi '(\phi
)^3 }{V(\phi ) \xi '(\phi ) \left(288 \alpha
^2 V(\phi )^2 \xi '(\phi )^2 V'(\phi )^2+8 V(\phi )^2
 \xi '(\phi )^3 (1-24 \alpha V(\phi ))
V'(\phi )+V(\phi )^2 \xi '(\phi )^2 \left(24 \alpha V(\phi )+32
V(\phi )^2 \xi '(\phi )^2+3\right)\right)}}}{\sqrt{3}}\, ,
\end{align}
where we took into account relation (\ref{Vdif1}). As it is
obvious from the expression (\ref{wavespeedanalyticalexpr}), the
analytic treatment of $c_A$ is impossible, since there is no
restriction constraining directly $\xi '$ and the scalar potential
and its derivative with respect to the scalar field, namely $V$
and $V'$. Thus only a numerical study is possible for the model
under study in order to reveal whether ghost instabilities occur
in the perturbations. The clue point of our analysis was that we
used the values for the free parameters that guaranteed the
inflationary phenomenological viability of the model. For example
in Fig. 4 we presented the behavior of the wave speed
$c_A$ for the values of the free parameters chosen in such a way
so that the inflationary viability of the model (\ref{xiB}) is
guaranteed. Actually, the values of the free parameters were
chosen to be exactly the ones which rendered the model (\ref{xiB})
viable with regard to its inflationary phenomenology.

We also need to discuss why the wave speed $c_A$ determines
whether ghost instabilities occur. This issue can be understood if
we look at equation (\ref{bigdiffeeqannew}), where $c_A$ is
interpreted as the wave speed of the scalar perturbation
$\Phi=\varphi_{\delta \phi}$, however we need to further clarify
this, since there is also another scalar component which is
perturbed in the Einstein-Gauss-Bonnet model at hand, and also we
need to discuss from which perturbed action these perturbations
originate. We shall use the notation of \cite{Hwang:2005hb}. The
other scalar perturbation associated with the theoretical
framework at hand is $\Psi$ defined as,
\begin{equation}\label{psidef}
\Psi=\varphi_{\chi}+\frac{\dot{F}+Q_a}{2F+Q_b}\frac{\delta
F_{\chi}}{\dot{F}}\, ,
\end{equation}
and the evolution of this perturbation mode is governed by the
following equation,
\begin{align}\label{bigevolutioneqn}
& \frac{\omega
\dot{\phi}^2+3\frac{(\dot{F}+Q_a)^2}{2F+Q_b}+Q_c+Q_d+\frac{\dot{F}+Q_a}{2F+Q_b}Q_e+\left(
\frac{\dot{F}+Q_a}{2F+Q_b}\right)^2Q_f}{\left(H\frac{\dot{F}+Q_a}{2F+Q_b}
\right)(F+\frac{1}{2}Q_b)}\times \\ \notag &
\frac{d}{dt}\Big{(}\frac{\left(H+\frac{\dot{F}+Q_a}{2F+Q_b}
\right)^2}{a\left(\omega
\dot{\phi}^2+3\frac{(\dot{F}+Q_a)^2}{2F+Q_b}+Q_c+Q_d+\frac{\dot{F}+Q_a}{2F+Q_b}Q_e+\left(\frac{\dot{F}+Q_a}{2F+Q_b}
\right)^2Q_f
\right)}\dot{\mathcal{S}}\Big{)}=c_A^2\frac{\Delta}{a^2}\Psi\, ,
\end{align}
where $\mathcal{S}$ is defined as follows,
\begin{equation}\label{mathcals}
\mathcal{S}=\frac{a(F+\frac{1}{2}Q_b)}{H+\frac{\dot{F}+Q_a}{2F+Q_b}}\Psi\,
.
\end{equation}
As it can be seen from both the wave equations
(\ref{bigdiffeeqannew}) and (\ref{bigevolutioneqn}), the quantity
$c_A^2$ plays the role of wave speed of both the fluctuating field
and of the perturbed metric. Since we assumed that the spacetime
has a flat spatial part, the sound wave speed
$c_s^2=\frac{\dot{p}}{\dot{\rho}}$ is identical with the sound
wave speed $c_A$. This feature can also be seen in Ref.
\cite{Hwang:2005hb} for the flat FRW case, in their table I, last
page of their article.

To further see that the quantity $c_A^2$ acts as a wave speed for
the evolution of the perturbed metric and for the field
perturbations, we shall bring Eqs. (\ref{bigdiffeeqannew}) and
(\ref{bigevolutioneqn}) to the more familiar Mukhanov-Sasaki form.
Following \cite{Hwang:2005hb}, we define $\bar{z}=c_Az$, where $z$
is defined as follows,
\begin{equation}\label{zetadef}
z=\frac{a\dot{\phi}}{(1+\epsilon_5)H}\sqrt{\frac{E}{F}}\, ,
\end{equation}
where $E$ is defined in Eq. (\ref{E1}) for the model at hand. Also
upon defining $\bar{v}=z\Phi$ and
$u=\frac{a}{\kappa^2H}\frac{1}{\bar{z}}\Psi$, the wave equations
(\ref{bigdiffeeqannew}) and (\ref{bigevolutioneqn}) can be cast to
the more familiar Mukhanov-Sasaki form, which is,
\begin{equation}\label{msform}
\bar{v}''-\left(c_A^2\Delta+\frac{z''}{z}\right)\bar{v}=0\, ,
\end{equation}
\begin{equation}\label{msform1}
u''-\left(c_A^2\Delta+\frac{(1/\bar{z})''}{1/\bar{z}}\right)u=0\,
,
\end{equation}
and clearly $c_A$ has the role of wave speed for both the
fluctuating fluid or field, and the perturbed metric.

Finally, let us note that the perturbed action for the fluctuating
field takes the form,
\begin{equation}\label{perturbedactionsec}
\delta^2S=\frac{1}{2}\int
az^2\left(\dot{\Phi}^2-c_A^2\frac{1}{a^2}\Phi_{,\mu}\Phi^{,\mu}
\right) dt d^3x\, ,
\end{equation}
or in terms of the conformal time, and the variable $\bar{v}$, the
above becomes,
\begin{equation}\label{perturbedactionsec1}
\delta^2S=\frac{1}{2}\int
\left(\bar{v}^2-c_A^2\bar{v}_{,\mu}\bar{v}^{,\mu}+\frac{z''}{z}\bar{v}^2
\right) d\eta d^3x\, .
\end{equation}
Thus it is apparent that for the flat spacetime case, the quantity
$c_A$ plays the role of the wave speed, which is identical to the
sound speed, only for the flat spacetime case. For more details on
this issue, we refer the reader to Ref. \cite{Hwang:2005hb}.

\section{Conclusions}

In this paper we studied a string and $R^2$ corrected canonical
minimally coupled scalar field theory and we demonstrated that the
$R^2$ gravity corrections do also provide a phenomenologically
viable theory. In this framework, by restoring the compatibility
of the theory with the GW170817 event, we showed that the scalar
potential and the Gauss-Bonnet scalar coupling functions are
interrelated via a specific differential equation, hence they
cannot be arbitrarily chosen, otherwise the continuity equation of
the scalar field is not satisfied. Furthermore, it becomes
apparent that some of the terms related to the string corrections,
are inferior compared to the rest terms in the equations of
motion, thus discarding them simplifies the equations greatly.
Doing so does not imply that their presence is also discarded as
string corrections, since these are also present in the kinetic
term due to the time derivative of the scalar field, a relation
which is derived from the realization that gravitational waves
propagate through spacetime with the speed of light. Moreover,
when the constant coupling $\alpha$ of the $R^2$ correction term,
satisfies $\alpha<10^{-3}$ in Planck units, or in other words, for
mass scales greater that $10^{20}$ GeV in natural units, the
contribution of the $R^2$ term becomes inferior compared to the
string corrections, and thus, such terms may be omitted from the
equations of motion as well. Although negligible, the presence of
the $R^2$ term alters the dynamics of the theory, since now the
time derivative of the term $\frac{\partial f}{\partial R}$ is
nonzero, thus the slow-roll index $\epsilon_3$ is non-zero. Also
the term $\frac{\partial f}{\partial R}$, for such choices of the
parameter $\alpha$, is close to, or equal to, unity in Planck
Units. As a last comment, it is worth stating that a different
choice of the Gauss-Bonnet scalar coupling function $\xi(\phi)$
may need in principle, different $\alpha$ values, in order for
this framework to produce a viable phenomenology. Also, the same
could be said about the case of having a non-minimally coupled
scalar theory with $R^2$ and string corrections, of the form
$f(R,\phi)=h(\phi)R+\alpha R^2$ gravity. We hope to address this
issue in a future work. Also, let us comment that the newly
introduced branch of astronomy, namely the gravitational wave
astronomy, can shed light on the issue of gravitational wave
speed, both for astrophysical and primordial gravitational waves,
and several interesting works have already appeared focusing on
this issue \cite{Nair:2019iur,Carson:2020cqb,Giare:2020vss}.

Finally, one issue that we briefly discussed is the possible
occurrence of ghost degrees of freedom in the context of the $R^2$
corrected Einstein-Gauss-Bonnet gravity. As we showed, at a
cosmological level, the perturbations of the Jordan frame theory
is free of ghosts, since quantitatively the speed of sound of the
scalar perturbations is equal to unity.


\begin{thebibliography}{99}








\bibitem{GBM:2017lvd}
  B.~P.~Abbott {\it et al.}
  ``Multi-messenger Observations of a Binary Neutron Star Merger,''
  Astrophys.\ J.\  {\bf 848} (2017) no.2,  L12
  doi:10.3847/2041-8213/aa91c9
  [arXiv:1710.05833 [astro-ph.HE]].




\bibitem{Ezquiaga:2017ekz}
  J.~M.~Ezquiaga and M.~Zumalacarregui,
  Phys.\ Rev.\ Lett.\  {\bf 119} (2017) no.25,  251304
  doi:10.1103/PhysRevLett.119.251304
  [arXiv:1710.05901 [astro-ph.CO]].










\bibitem{Hwang:2005hb}
  J.~c.~Hwang and H.~Noh,
  Phys.\ Rev.\ D {\bf 71} (2005) 063536
  doi:10.1103/PhysRevD.71.063536
  [gr-qc/0412126].


\bibitem{Nojiri:2006je}
  S.~Nojiri, S.~D.~Odintsov and M.~Sami,
  Phys.\ Rev.\ D {\bf 74} (2006) 046004
  doi:10.1103/PhysRevD.74.046004
  [hep-th/0605039].




\bibitem{Cognola:2006sp}
  G.~Cognola, E.~Elizalde, S.~Nojiri, S.~Odintsov and S.~Zerbini,
  Phys.\ Rev.\ D {\bf 75} (2007) 086002
  doi:10.1103/PhysRevD.75.086002
  [hep-th/0611198].



\bibitem{Nojiri:2005vv}
  S.~Nojiri, S.~D.~Odintsov and M.~Sasaki,
  Phys.\ Rev.\ D {\bf 71} (2005) 123509
  doi:10.1103/PhysRevD.71.123509
  [hep-th/0504052].


\bibitem{Nojiri:2005jg}
  S.~Nojiri and S.~D.~Odintsov,
  Phys.\ Lett.\ B {\bf 631} (2005) 1
  doi:10.1016/j.physletb.2005.10.010
  [hep-th/0508049].







\bibitem{Satoh:2007gn}
  M.~Satoh, S.~Kanno and J.~Soda,
  Phys.\ Rev.\ D {\bf 77} (2008) 023526
  doi:10.1103/PhysRevD.77.023526
  [arXiv:0706.3585 [astro-ph]].



\bibitem{Bamba:2014zoa}
  K.~Bamba, A.~N.~Makarenko, A.~N.~Myagky and S.~D.~Odintsov,
  JCAP {\bf 1504} (2015) 001
  doi:10.1088/1475-7516/2015/04/001
  [arXiv:1411.3852 [hep-th]].


\bibitem{Yi:2018gse}
  Z.~Yi, Y.~Gong and M.~Sabir,
  Phys.\ Rev.\ D {\bf 98} (2018) no.8,  083521
  doi:10.1103/PhysRevD.98.083521
  [arXiv:1804.09116 [gr-qc]].


\bibitem{Guo:2009uk}
  Z.~K.~Guo and D.~J.~Schwarz,
  Phys.\ Rev.\ D {\bf 80} (2009) 063523
  doi:10.1103/PhysRevD.80.063523
  [arXiv:0907.0427 [hep-th]].


\bibitem{Guo:2010jr}
  Z.~K.~Guo and D.~J.~Schwarz,
  Phys.\ Rev.\ D {\bf 81} (2010) 123520
  doi:10.1103/PhysRevD.81.123520
  [arXiv:1001.1897 [hep-th]].


\bibitem{Jiang:2013gza}
  P.~X.~Jiang, J.~W.~Hu and Z.~K.~Guo,
  Phys.\ Rev.\ D {\bf 88} (2013) 123508
  doi:10.1103/PhysRevD.88.123508
  [arXiv:1310.5579 [hep-th]].



\bibitem{Kanti:2015pda}
  P.~Kanti, R.~Gannouji and N.~Dadhich,
  Phys.\ Rev.\ D {\bf 92} (2015) no.4,  041302
  doi:10.1103/PhysRevD.92.041302
  [arXiv:1503.01579 [hep-th]].


\bibitem{vandeBruck:2017voa}
  C.~van de Bruck, K.~Dimopoulos, C.~Longden and C.~Owen,
  arXiv:1707.06839 [astro-ph.CO].



\bibitem{Kanti:1998jd}
  P.~Kanti, J.~Rizos and K.~Tamvakis,
  Phys.\ Rev.\ D {\bf 59} (1999) 083512
  doi:10.1103/PhysRevD.59.083512
  [gr-qc/9806085].




\bibitem{Pozdeeva:2020apf}
  E.~O.~Pozdeeva, M.~R.~Gangopadhyay, M.~Sami, A.~V.~Toporensky and S.~Y.~Vernov,
  arXiv:2006.08027 [gr-qc].

\bibitem{Fomin:2020hfh}
  I.~Fomin,
  arXiv:2004.08065 [gr-qc].

\bibitem{DeLaurentis:2015fea}
  M.~De Laurentis, M.~Paolella and S.~Capozziello,
  Phys.\ Rev.\ D {\bf 91} (2015) no.8,  083531
  doi:10.1103/PhysRevD.91.083531
  [arXiv:1503.04659 [gr-qc]].


\bibitem{Chervon:2019sey}
  S.~Chervon, I.~Fomin, V.~Yurov and A.~Yurov,
  doi:10.1142/11405



\bibitem{Nozari:2017rta}
  K.~Nozari and N.~Rashidi,
  Phys.\ Rev.\ D {\bf 95} (2017) no.12,  123518
  doi:10.1103/PhysRevD.95.123518
  [arXiv:1705.02617 [astro-ph.CO]].




\bibitem{Odintsov:2018zhw}
  S.~D.~Odintsov and V.~K.~Oikonomou,
  Phys.\ Rev.\ D {\bf 98} (2018) no.4,  044039
  doi:10.1103/PhysRevD.98.044039
  [arXiv:1808.05045 [gr-qc]].


  \bibitem{Kawai:1998ab}
  S.~Kawai, M.~a.~Sakagami and J.~Soda,
  Phys.\ Lett.\ B {\bf 437}, 284 (1998)
  doi:10.1016/S0370-2693(98)00925-3
  [gr-qc/9802033].


\bibitem{Yi:2018dhl}
  Z.~Yi and Y.~Gong,
  Universe {\bf 5} (2019) no.9,  200
  doi:10.3390/universe5090200
  [arXiv:1811.01625 [gr-qc]].


\bibitem{vandeBruck:2016xvt}
  C.~van de Bruck, K.~Dimopoulos and C.~Longden,
  Phys.\ Rev.\ D {\bf 94} (2016) no.2,  023506
  doi:10.1103/PhysRevD.94.023506
  [arXiv:1605.06350 [astro-ph.CO]].


\bibitem{Kleihaus:2019rbg}
  B.~Kleihaus, J.~Kunz and P.~Kanti,
  arXiv:1910.02121 [gr-qc].





\bibitem{Bakopoulos:2019tvc}
  A.~Bakopoulos, P.~Kanti and N.~Pappas,
  Phys.\ Rev.\ D {\bf 101} (2020) no.4,  044026
  doi:10.1103/PhysRevD.101.044026
  [arXiv:1910.14637 [hep-th]].


\bibitem{Maeda:2011zn}
  K.~i.~Maeda, N.~Ohta and R.~Wakebe,
  Eur.\ Phys.\ J.\ C {\bf 72} (2012) 1949
  doi:10.1140/epjc/s10052-012-1949-6
  [arXiv:1111.3251 [hep-th]].






\bibitem{Bakopoulos:2020dfg}
  A.~Bakopoulos, P.~Kanti and N.~Pappas,
  arXiv:2003.02473 [hep-th].


\bibitem{Ai:2020peo}
W.~Ai,
[arXiv:2004.02858 [gr-qc]].



\bibitem{Odintsov:2019clh}
  S.~D.~Odintsov and V.~K.~Oikonomou,
  Phys.\ Lett.\ B {\bf 797} (2019) 134874
  doi:10.1016/j.physletb.2019.134874
  [arXiv:1908.07555 [gr-qc]].



\bibitem{Oikonomou:2020oil}
V.~K.~Oikonomou and F.~P.~Fronimos,
[arXiv:2007.11915 [gr-qc]].

\bibitem{Odintsov:2020xji}
S.~D.~Odintsov, V.~K.~Oikonomou and F.~P.~Fronimos,
Annals Phys. \textbf{420} (2020), 168250
doi:10.1016/j.aop.2020.168250 [arXiv:2007.02309 [gr-qc]].



\bibitem{Oikonomou:2020sij}
V.~K.~Oikonomou and F.~P.~Fronimos,
[arXiv:2006.05512 [gr-qc]].



\bibitem{Odintsov:2020zkl}
S.~D.~Odintsov and V.~K.~Oikonomou,
Phys. Lett. B \textbf{805} (2020), 135437
doi:10.1016/j.physletb.2020.135437 [arXiv:2004.00479 [gr-qc]].


\bibitem{Odintsov:2020sqy}
S.~D.~Odintsov, V.~K.~Oikonomou and F.~P.~Fronimos,
[arXiv:2003.13724 [gr-qc]].





\bibitem{Easther:1996yd}
  R.~Easther and K.~i.~Maeda,
  Phys.\ Rev.\ D {\bf 54} (1996) 7252
  doi:10.1103/PhysRevD.54.7252
  [hep-th/9605173].

\bibitem{Antoniadis:1993jc}
  I.~Antoniadis, J.~Rizos and K.~Tamvakis,
  Nucl.\ Phys.\ B {\bf 415} (1994) 497
  doi:10.1016/0550-3213(94)90120-1
  [hep-th/9305025].

\bibitem{Antoniadis:1990uu}
I.~Antoniadis, C.~Bachas, J.~R.~Ellis and D.~V.~Nanopoulos,
Phys.\ Lett.\ B \textbf{257} (1991), 278-284
doi:10.1016/0370-2693(91)91893-Z




\bibitem{Kanti:1995vq}
P.~Kanti, N.~Mavromatos, J.~Rizos, K.~Tamvakis and E.~Winstanley,
Phys. Rev. D \textbf{54} (1996), 5049-5058
doi:10.1103/PhysRevD.54.5049 [arXiv:hep-th/9511071 [hep-th]].



\bibitem{Kanti:1997br}
P.~Kanti, N.~Mavromatos, J.~Rizos, K.~Tamvakis and E.~Winstanley,
Phys. Rev. D \textbf{57} (1998), 6255-6264
doi:10.1103/PhysRevD.57.6255 [arXiv:hep-th/9703192 [hep-th]].




\bibitem{Nojiri:2017ncd}
S.~Nojiri, S.~D.~Odintsov and V.~K.~Oikonomou,
Phys.\ Rept.\ {\bf 692} (2017) 1 doi:10.1016/j.physrep.2017.06.001
[arXiv:1705.11098 [gr-qc]].

\bibitem{Nojiri:2010wj}
S.~Nojiri and S.~D.~Odintsov,
Phys.\ Rept.\ {\bf 505} (2011) 59
doi:10.1016/j.physrep.2011.04.001 [arXiv:1011.0544 [gr-qc]].





\bibitem{Nojiri:2006ri}
S.~Nojiri and S.~D.~Odintsov,
eConf C {\bf 0602061} (2006) 06
 [Int.\ J.\ Geom.\ Meth.\ Mod.\ Phys.\ {\bf 4} (2007) 115]
doi:10.1142/S0219887807001928 [hep-th/0601213].

\bibitem{Capozziello:2011et}
S.~Capozziello and M.~De Laurentis,
Phys.\ Rept.\ {\bf 509} (2011) 167
doi:10.1016/j.physrep.2011.09.003 [arXiv:1108.6266 [gr-qc]].

\bibitem{Capozziello:2010zz}
V.~Faraoni and S.~Capozziello,
Fundam.\ Theor.\ Phys.\ {\bf 170} (2010).
doi:10.1007/978-94-007-0165-6


\bibitem{Olmo:2011uz}
G.~J.~Olmo,
Int.\ J.\ Mod.\ Phys.\ D {\bf 20} (2011) 413
doi:10.1142/S0218271811018925 [arXiv:1101.3864 [gr-qc]].


\bibitem{Nojiri:2003ft}
S.~Nojiri and S.~D.~Odintsov,
Phys. Rev. D \textbf{68} (2003), 123512
doi:10.1103/PhysRevD.68.123512 [arXiv:hep-th/0307288 [hep-th]].



\bibitem{Sa:2020qfd}
P.~M.~Sa,
Universe \textbf{6} (2020) no.6, 78 [arXiv:2002.09446 [gr-qc]].








\bibitem{Akrami:2018odb}
  Y.~Akrami {\it et al.} [Planck Collaboration],
  arXiv:1807.06211 [astro-ph.CO].





\bibitem{Oikonomou:2020krq}
V.~K.~Oikonomou,
EPL \textbf{130} (2020) no.1, 10006
doi:10.1209/0295-5075/130/10006 [arXiv:2004.10778 [gr-qc]].




\bibitem{Nair:2019iur}
R.~Nair, S.~Perkins, H.~O.~Silva and N.~Yunes,
Phys. Rev. Lett. \textbf{123} (2019) no.19, 191101
doi:10.1103/PhysRevLett.123.191101 [arXiv:1905.00870 [gr-qc]].


\bibitem{Carson:2020cqb}
Z.~Carson and K.~Yagi,
[arXiv:2002.08559 [gr-qc]].



\bibitem{Giare:2020vss}
W.~Giare and F.~Renzi,
[arXiv:2007.04256 [astro-ph.CO]].





\end{thebibliography}
\end{document}